\documentclass[aps,prl,twocolumn,superscriptaddress,longbibliography,nofootinbib]{revtex4-2}
\usepackage{amsmath,amssymb,amsthm,bm}
\usepackage{graphicx}
\usepackage{tikz}
\usepackage{hyperref}
\hypersetup{colorlinks=true,linkcolor=blue!60!black,citecolor=blue!60!black,urlcolor=blue!60!black}

\newtheorem{lemma}{Lemma}
\newtheorem{proposition}{Proposition}

\newcommand{\bS}{\mathbf S}
\newcommand{\bh}{\mathbf h}
\newcommand{\Jt}{\mathsf J}
\newcommand{\Tt}{\mathsf T}
\newcommand{\he}{\mathfrak h}
\newcommand{\tr}{\operatorname{Tr}}
\newcommand{\ket}[1]{|#1\rangle}
\newcommand{\bra}[1]{\langle#1|}

\begin{document}

\title{Exact many-body scars of generic spin models from singlet covers}

\author{Camilla Polvara}
\affiliation{D\'epartement de Physique, Universit\'e de Montr\'eal, Montr\'eal, QC H3C 3J7, Canada}
\author{William Witczak-Krempa}
\affiliation{D\'epartement de Physique, Centre de Recherches Math\'ematiques, and Institut Courtois, Universit\'e de Montr\'eal, Montr\'eal, QC H3C 3J7, Canada}

\date{\today}

\begin{abstract}
The square-lattice transverse-field Ising model has exact zero-energy eigenstates at any field: products of singlets on alternate plaquette diagonals. A longitudinal field isolates them from a symmetry-protected degeneracy. The mechanism is that a site coupled equally to both ends of a singlet annihilates it. The result thus holds for any spin s, anisotropic exchange, a large class of bond disorder and certain three-spin terms. This mechanism decides which lattices host such states: two of the 11 Archimedean lattices (four at s=1/2), bcc and quasicrystals with spins on the edges. Singlet pumping cools any state into them.
\end{abstract}

\maketitle
\newcommand{\Wt}{\mathsf W}

\begin{figure}[t]
\centering
\resizebox{\columnwidth}{!}{%
\begin{tikzpicture}[scale=0.62]
\begin{scope}[scale=1.18]
\foreach \x in {0,...,5}{\draw[black!25] (\x,-0.5)--(\x,5.5);\draw[black!25] (-0.5,\x)--(5.5,\x);}
\foreach \x in {0,...,5}\foreach \y in {0,...,5}{\fill[black!70] (\x,\y) circle (0.09);}
\foreach \x in {0,2,4}\foreach \y in {0,2,4}{\draw[line width=2.2pt,blue!70!black] (\x,\y)--(\x+1,\y+1);}
\foreach \x in {1,3}\foreach \y in {1,3}{\draw[line width=2.2pt,red!75!black] (\x+1,\y)--(\x,\y+1);}
\foreach \y in {1,3}{\draw[line width=2.2pt,red!75!black,dashed] (5,\y+1)--(5.5,\y+0.5);\draw[line width=2.2pt,red!75!black,dashed] (0,\y)--(-0.5,\y+0.5);}
\foreach \x in {1,3}{\draw[line width=2.2pt,red!75!black,dashed] (\x+1,5)--(\x+0.5,5.5);\draw[line width=2.2pt,red!75!black,dashed] (\x,0)--(\x+0.5,-0.5);}
\draw[line width=3.2pt,orange,opacity=0.85] (3,2)--(2,2);\draw[line width=3.2pt,orange,opacity=0.85] (3,2)--(3,3);
\fill[orange] (3,2) circle (0.14);
\fill[green!55!black,opacity=0.35] (0,4)--(1,4)--(0,5)--cycle;
\fill[violet,opacity=0.35] (1,5)--(1,4)--(0,5)--cycle;
\node[font=\bfseries\footnotesize,anchor=south west] at (-0.6,5.55) {(a)};
\end{scope}
\node[anchor=south west,inner sep=0] at (6.9,-0.75) {\includegraphics[width=4.3cm]{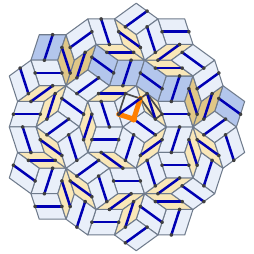}};
\node[font=\bfseries\footnotesize,anchor=south west] at (6.9,6.549) {(b)};
\end{tikzpicture}}
\caption{(a) The state $\ket{M_{00}}$: singlets on the ``/'' diagonals of even--even plaquettes (blue) and the ``\textbackslash'' diagonals of odd--odd plaquettes (red); dashed segments continue across the periodic boundary. The orange wedge, a site coupled equally to both ends of a singlet pair (a dimer), annihilates the state. The two shaded three-spin terms, at the two ends of a dimer and sharing its two common neighbours, cancel on the state (Proposition~\ref{prop:multi}). (b) Spins on the edges of a Penrose tiling, coupled when their edges are adjacent sides of a tile (grey bonds, shown for two tiles). One singlet (blue) crosses every tile between two opposite edges, alternating along every ribbon of tiles (one is shaded); the opposite phase gives the second cover.}
\label{fig:covers}
\end{figure}

\emph{Introduction.}---The transverse-field Ising model on the square lattice is the textbook nonintegrable quantum magnet and the workhorse of two-dimensional thermalization studies \cite{Deutsch1991,Srednicki1994,Rigol2008,Chiba2024}. We show that, for either sign of the coupling and at every value of the field, it has four exact eigenstates in the middle of its spectrum: the product of singlets on the diagonals of alternate plaquettes [Fig.~\ref{fig:covers}(a)] and its three translates, at energy zero and with area-law entanglement, locally indistinguishable from infinite temperature except on their singlet pairs. The same states remain exact when a longitudinal or a staggered field is added, for Heisenberg, XXZ, XYZ and Dzyaloshinskii--Moriya exchange, for any spin, under half of all bond disorder, under any time dependence within this family and, for spin $\tfrac12$, under a class of three-spin interactions. They had gone unnoticed in the Ising model for three reasons: at zero longitudinal field they hide inside a symmetry-protected, exponentially degenerate zero-energy space \cite{Schecter2018}; they are SU(2) singlets in a model with no SU(2); and no single term of the Hamiltonian annihilates them.

The total spin of a pair annihilates its singlet, so a site coupled equally to both ends of a singlet pair, or dimer, annihilates the state: a \emph{wedge}, with that site as apex and the two bonds as legs. On the square lattice the $N$ wedges partition the $2N$ bonds: the uniform Hamiltonian is itself a sum of wedges. A wedge has the Shiraishi--Mori form $PhP$ \cite{Shiraishi2017}, with $P$ the non-singlet projector of a pair that is not a bond: the transverse-field Ising model is a Shiraishi--Mori Hamiltonian in disguise. For uniform couplings the condition for such a state involves only which sites are bonded, and a counting identity decides which lattices carry one. Four of the 11 Archimedean lattices do (the tilings of the plane by regular polygons with all vertices equivalent, such as the honeycomb and kagome lattices \cite{Richter2004}), as do bcc and the Penrose tiling once the spins sit on its edges [Fig.~\ref{fig:covers}(b)]; no other common lattice carries one, whatever the range of the singlets.

\begin{figure*}[t]
\centering
\includegraphics[width=\textwidth]{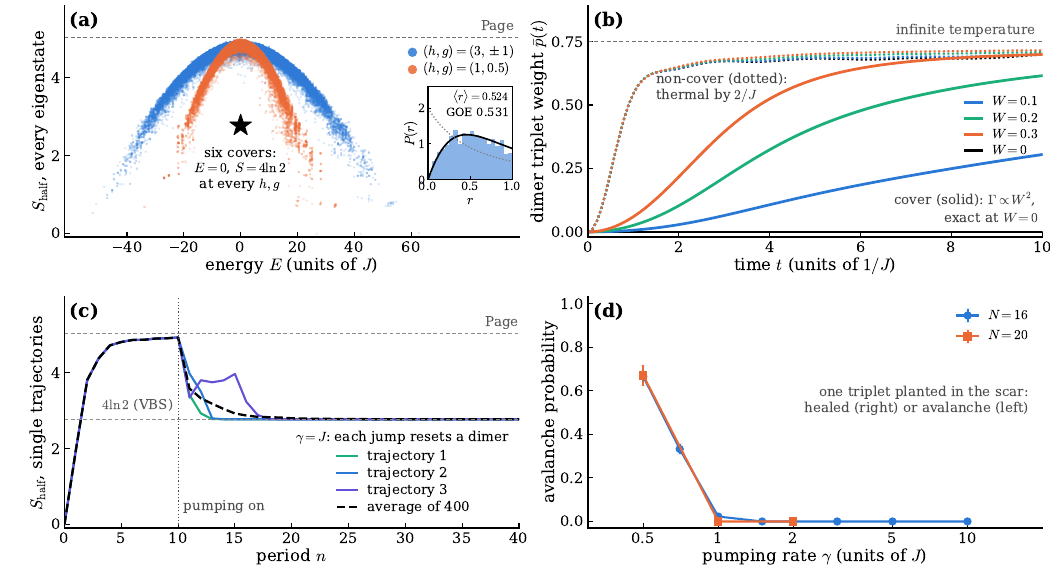}
\caption{Scar status of the covers, and cooling into them. (a) Half-torus entanglement entropy of all eigenstates of the $4\times4$ mixed-field Ising torus at $(h,g)=(3,\pm1)$ (blue) and $(1,0.5)$ (orange): the covers (star, $E=0$, $S=4\ln2$) span the zero-energy space, every other eigenstate near $E=0$ lies on the thermal arc peaking at the Page value (dashed). Inset: level-spacing ratios in the irreducible symmetry blocks \cite{SM}, with the GOE surmise (solid) and the Poisson law (dotted). (b) Heisenberg antiferromagnet $\sum J_{ij}\bS_i\cdot\bS_j+\sum h_iS^z_i$ on a 20-site torus, with every bond $J_{ij}=J(1+W\epsilon_{ij})$ and every field $h_i=WJ\eta_i$ random ($\epsilon,\eta$ uniform in $[-1,1]$), quenched from the cover (solid) and from a product of diagonal singlets that is not a cover (dotted): mean triplet weight of their own dimers over 100 realizations. (c) Periodic switch between the two models of (a) on the $4\times4$ torus, from the N\'eel state, with pumping at $\gamma=J$ from period 10: three single trajectories and the average of 400. (d) Probability that one triplet planted in the scar spreads to at least half of the dimers, with binomial errors.}
\label{fig:drive}
\end{figure*}

No Hamiltonian of the family acts on the state, so pumping each pair into its singlet makes it the unique attractor under any of them and cools the system into it from any initial state [Fig.~\ref{fig:drive}(c,d)].

Products of singlets are the oldest exact eigenstates of spin models \cite{Majumdar1969,Shastry1981,Kumar2002}, and their embedding by pair projectors underlies the Shastry--Sutherland and maple-leaf scars \cite{McClarty2020}. Petrovich, Rossignoli and Canosa \cite{Petrovich2024} characterized the quadratic Hamiltonians sharing a given singlet product; Dai \cite{Dai2024} found the diagonal valence-bond solids (VBS) as zero-energy eigenstates of the square-lattice Heisenberg antiferromagnet, and van Voorden \emph{et al.} \cite{vanVoorden2020} their analogue in the transverse-field Ising ladder. Related constructions are the graph rules for product eigenstates of XYZ models \cite{Bhowmick2025} and the matrix-product eigenstates of specific non-frustration-free Hamiltonians, built from local errors that cancel globally \cite{Gehrmann2026,GarreRubio2026}. Next to these, and to the exact scars of other two-dimensional \cite{Lin2020,Miao2026,Hartse2024} and one-dimensional models \cite{Moudgalya2018a,Moudgalya2018b,Schecter2019,Turner2018,Lin2019,Yang1989,Mark2020,Moudgalya2020,Pakrouski2020,Ren2021,Sala2020,Khemani2020,Moudgalya2022}, what is new is that the criterion classifies lattices rather than Hamiltonians, and that it applies to the standard models of quantum chaos.

\emph{The state and its proof.}---Let $G$ be a graph with $N$ sites and bond weights $J_{ij}$, uniform unless stated, with a spin $s$ on every site, and consider the whole family
\begin{equation}
H(\Jt,\bh)=\sum_{(ij)}J_{ij}\,\bS_i^{\mathsf T}\Jt\,\bS_j+\bh\cdot\bS_{\rm tot},
\label{eq:family}
\end{equation}
the sum over bonds, oriented from sublattice A to B on a bipartite lattice, with $\Jt$ any real $3\times3$ tensor and $\bh$ any field. The symmetric part of $\Jt$ gives Ising, XXZ, XYZ and Heisenberg exchange, the antisymmetric part a uniform Dzyaloshinskii--Moriya (DM) term. For $s=\tfrac12$, $\Jt={\rm diag}(0,0,4)$ and $\bh=(2h,0,2g)$ this is the mixed-field Ising model $\sum J_{ij}\sigma^z_i\sigma^z_j+h\sum\sigma^x_i+g\sum\sigma^z_i$. A pairing $M$ of all sites into dimers $d=(a,a')$, which need not be bonds, defines the product $\ket M_s$ of their singlets. This state is an SU(2) singlet, and the total spin $\bS^+_d=\bS_a+\bS_{a'}$ of any dimer annihilates it. The pairing is a \emph{singlet cover} when the bonds between different dimers can be grouped into wedges. On the $L_x\times L_y$ square lattice with periodic boundaries (a torus) and even sides, label each plaquette by its lower-left corner $(x,y)$ and put a singlet on its ``/'' diagonal if $x$ and $y$ are both even, on its ``\textbackslash'' diagonal if both are odd [Fig.~\ref{fig:covers}(a)]. Translations give four such states $\ket{M_{\mu\nu}}$, the Shastry--Sutherland dimer pattern \cite{Shastry1981} without its diagonal coupling.

\begin{proposition}\label{prop:vbs}
For every even $L_x,L_y\ge4$ and every spin $s$, the four states $\ket{M_{\mu\nu}}_s$ are exact zero-energy eigenstates of $H(\Jt,\bh)$ for every tensor $\Jt$, DM included, and every field.
\end{proposition}

\begin{proof}
Every bond is a leg of exactly one wedge. For $a$ in the ``/'' dimer $(a,a+\hat x+\hat y)$, the bonds from $a$ to $a+\hat x$ and $a+\hat y$ are legs of the two wedges on that dimer, and the bonds to $a-\hat x$ and $a-\hat y$ are the two legs of the wedge with apex $a$ on the ``\textbackslash'' dimer $(a-\hat x,a-\hat y)$; likewise at every other site. The $N$ wedges partition the $2N$ bonds, and the A$\to$B orientation is \emph{consistent}: the two legs of every wedge point both towards its apex or both away from it, as DM requires. With uniform couplings $H$ is thus $\bh\cdot\bS_{\rm tot}$ plus one term per wedge, $\bS_c^{\mathsf T}\Jt(\bS_a+\bS_{a'})$ for an apex $c$ on A and $(\bS_a+\bS_{a'})^{\mathsf T}\Jt\,\bS_c$ on B, and every term annihilates $\ket{M_{\mu\nu}}_s$.
\end{proof}

The field may differ from dimer to dimer: it need only be equal at the two ends of each. A N\'eel-staggered field qualifies, since a diagonal joins two sites of one sublattice. The energy $E=0=\tr H/(2s+1)^N$ is the infinite-temperature energy. The entanglement is area law, $S_A=n_{\rm cut}\ln(2s+1)$ for a cut crossing $n_{\rm cut}$ dimers, and all nearest-neighbour correlators vanish, while $\langle\bS_a\cdot\bS_{a'}\rangle=-s(s+1)$ on every dimer. The state is a scar in the strict sense, an isolated area-law eigenstate at the centre of a chaotic spectrum [Fig.~\ref{fig:drive}(a)]: on the $4\times4$ torus the level statistics are those of the GOE, $\langle r\rangle=0.524\pm0.006$ against $0.531$, every other eigenstate near $E=0$ lies on the thermal arc, and the zero-energy space is exactly the span of the covers at every field examined, as on a 20-site torus \cite{SM}. At $g=0$ the covers hide inside the $770$-fold degenerate zero-energy space that the spectral reflection symmetry protects \cite{Schecter2018}, conjectured to contain area-law states \cite{Karle2021,Ivanov2025}; the longitudinal field, which annihilates them, splits the rest at order $g^2$ \cite{SM}.

The proof generalizes to any graph: the Ising part $H^{zz}=\sum J_{ij}S^z_iS^z_j$ is diagonal, and $\ket M_s$ lives on the configurations in which every dimer is antiparallel, where its value is $-\sum_dJ_{aa'}m_d^2+\sum_{d<d'}W_{dd'}m_dm_{d'}$ in the dimer moments $m_d$, with $W_{dd'}=J_{ab}-J_{ab'}-J_{a'b}+J_{a'b'}$ the signed count of the bonds between $d=(a,a')$ and $d'=(b,b')$. So $\ket M_s$ is an eigenstate of $H^{zz}$ iff every $W_{dd'}$ vanishes (and, for $s\ge1$, no dimer is a bond): iff the classical Ising energy is the same on all antiparallel configurations. Being a singlet, it is then an eigenstate of $H(\Jt,\bh)$ for every symmetric $\Jt$ and every $\bh$, and for every $\Jt$ when no dimer is a bond and the bonds are consistently oriented. The condition involves the dimers in pairs only, so the state stays exact on any finite cluster made of whole dimers, with open boundaries. This is the balance condition of Ref.~\cite{Petrovich2024}; conversely, every two-body Hamiltonian with $\ket M_s$ as an eigenstate is a sum of wedges $\mathbf X_d\cdot\bS^+_d$ and intra-dimer terms (Theorem 2 of Ref.~\cite{Petrovich2024}; short proofs in the SM \cite{SM}).

\emph{Which lattices carry such a state.}---With uniform couplings $W_{dd'}=0$ is the rule that defines a cover: the bonds between two dimers form nothing, one wedge, or two wedges, that is all four bonds, a square with the two dimers as its diagonals. Call a site that neighbours both ends of a dimer a \emph{middle} of it, and let $|C(d)|$ be the number of middles of $d$.

\begin{proposition}[classification]\label{prop:count}
For a cover with $B$ dimers on bonds and $K$ pairs of dimers joined by all four bonds, on a lattice with $\mathcal E$ bonds,
\begin{equation}
\textstyle\sum_{d\in M}|C(d)|=\tfrac12(\mathcal E-B)+2K ,
\label{eq:count}
\end{equation}
so the dimers have on average at least half as many middles as a site has neighbours other than its partner, whereas in a planar lattice whose bonds all have the same length a pair of sites has at most two common neighbours. Consequently: (i) on the square torus with even sides $L_x,L_y\ge4$ the covers, with dimers of any range, are exactly the four diagonal VBS, with two more on $4\times4$, and there are none if a side is odd; (ii) exactly four of the 11 Archimedean lattices admit covers: the square lattice and the ruby lattice of triangles, squares and hexagons, with dimers off the bonds, and the Shastry--Sutherland and maple-leaf lattices, whose only covers are their known on-bond dimer states \cite{Shastry1981,McClarty2020}; (iii) of the simple cubic, bcc, fcc, hcp, diamond and pyrochlore lattices, only bcc admits covers; (iv) a tiling of the plane by rhombi, with a spin on every vertex, admits a cover only if almost all vertices have four neighbours, which excludes the Penrose and Ammann--Beenker quasicrystals and the dice lattice.
\end{proposition}

The identity \eqref{eq:count} counts neighbours: every neighbour of a site, other than its partner, is either a middle of the site's own dimer or belongs to a dimer of which the site is a middle \cite{SM}. On most lattices the resulting lower bound exceeds the geometric upper bound and there is no cover. Where the two meet, every dimer must have the maximal number of middles and a local rule propagates: on the square lattice every dimer is a plaquette diagonal and the partner of one site fixes the whole cover. Off the bonds (square, ruby, bcc, and the quasicrystals below) the states sit at $E=0$ for every spin and every tensor, DM included once the bonds are consistently oriented \cite{SM}; on the bonds (Shastry--Sutherland, maple-leaf) they are exact for $s=\tfrac12$ only, at energy $-\tfrac14\tr\Jt$ per dimer, and DM breaks them. The same holds on the icosahedron, where such states were first seen \cite{PolvaraWKfrh} in rank diagnostics of exact scars \cite{Parez2025,Yao2025}: its 30 edges split into five covers of six singlets each \cite{SM}. A rhombus tiling has four neighbours per site on average, so \eqref{eq:count} forces the dimers onto rhombus diagonals and every site to have an even number of neighbours, hence four, up to a vanishing fraction of the sites, whereas a finite fraction of the Penrose vertices have three. Moving the spins onto the edges repairs this [Fig.~\ref{fig:covers}(b)].

\begin{proposition}[quasiperiodic covers]\label{prop:penrose}
Consider a rhombus tiling built by de Bruijn's grid method \cite{deBruijn1981}, as the Penrose and Ammann--Beenker tilings are: it has one tile per crossing of a grid of several families of parallel lines, no three through one point, and the tiles along a line form a \emph{ribbon}, a chain of tiles sharing parallel edges. Put spins on the edges of the tiling, coupled when their edges are adjacent sides of a tile. One singlet per tile, between the midpoints of two opposite edges, with the two choices alternating along every ribbon, is a singlet cover; such an alternation exists for every such tiling and is unique up to the exchange of the two phases.
\end{proposition}

Each dimer has exactly two middles, the other two edges of its tile, and each site is the middle of exactly one dimer, so the counting rule is met. The alternation is consistent: its only possible conflict, around three mutually crossing ribbons, depends on the parity of the number of other ribbons crossing their triangle, which is even \cite{SM}. Each such cover is a zero-energy eigenstate of the Ising model on its lattice at every field.

\emph{Beyond two-body interactions.}---A term coupling a site to two of its neighbours meets three dimers, beyond the reach of any two-body criterion \cite{Petrovich2024}. The same argument applies: the state is an eigenstate of an Ising-type term of any order iff the signed count over the terms meeting each set of dimers an odd number of times vanishes \cite{SM}. Consider the three-spin interaction of every site with every pair of its neighbours,
\begin{equation}
H_3=\sum_{\bm r}\sum_{\{n,n'\}\subset N(\bm r)}\Tt_3^{\alpha\beta\gamma}\,S^\alpha_{\bm r}S^\beta_nS^\gamma_{n'},
\label{eq:multi}
\end{equation}
$N(\bm r)$ the four neighbours of $\bm r$: the same coupling on every path of two bonds, corner and straight alike. The tensor $\Tt_3$ may be any real tensor symmetric in its last two indices, those of the two neighbours: the three-spin Ising interaction $S^z_{\bm r}S^z_nS^z_{n'}$ and its rotations, but also $S^x_{\bm r}S^z_nS^z_{n'}$ or $(\bm v\cdot\bS_{\bm r})(\bS_n\cdot\bS_{n'})$; the scalar chirality $\bS_{\bm r}\cdot(\bS_n\times\bS_{n'})$, being antisymmetric, is excluded.

\begin{proposition}[three-spin interactions]\label{prop:multi}
For $s=\tfrac12$ the states $\ket{M_{\mu\nu}}$ are exact zero-energy eigenstates of $H_3$ for every $\Tt_3$ symmetric in its last two indices, hence of $H(\Jt,\bh)+H_3$.
\end{proposition}

The proof is the pairing of Fig.~\ref{fig:covers}(a): the six triples at a site combine into terms that contain the total spin of a dimer. The four neighbours of a site are the two middles of its own dimer and the two ends of a dimer of which it is a middle. Four triples pair among themselves through that dimer; the triple through the two middles pairs with that of the site's partner, which has the same middles; and the triple on that dimer reduces, for $s=\tfrac12$, to a field on the site, which pairs with the same field on its partner \cite{SM}. Each pair at one site contains a straight and a corner triple, so neither kind alone is admissible, nor are the plaquette four-spin term and further-neighbour exchange, while the five-spin star of a site and its four neighbours is \cite{SM}. The same pairing applies to the covers of the ruby and bcc lattices and of the quasicrystals; for $s\ge1$ it survives for $(\bm v\cdot\bS_{\bm r})(\bS_n\cdot\bS_{n'})$, a triple with two sites on one dimer otherwise exciting that dimer to spin two.

\emph{Half of all disorder is harmless.}---Uniform couplings were never required: the Hamiltonians of wedge form on the square lattice are those with equal tensors on the two legs of every wedge and fields constant on each dimer. Bond disorder thus splits into a harmless half, constant on each wedge, and the harmful leg imbalances $\delta J_{ca}-\delta J_{ca'}$, which destroy the Shastry--Sutherland scars \cite{McClarty2020}. For arbitrary disorder, bond tensors $\delta\Jt_{ij}$ and site fields $\delta\bh_i$, the triplet weight of a dimer grows as $\Gamma_dt^2$ with $\Gamma_d=\tfrac14|\delta\bh_a-\delta\bh_{a'}|^2+\tfrac1{16}\sum_{d'\ne d}\|\delta\Wt_{dd'}\|_F^2$, where $\delta\Wt_{dd'}=\delta\Jt_{ab}-\delta\Jt_{ab'}-\delta\Jt_{a'b}+\delta\Jt_{a'b'}$ \cite{SM}: only the field difference across the dimer and the tensor imbalances around it enter, so each dimer decays on a time $\sim\Gamma_d^{-1/2}$ that does not depend on the system size. Figure~\ref{fig:drive}(b) shows this for a Heisenberg magnet in which every bond and every field is random, next to a control: a second product of diagonal singlets, with the same energy and nearest-neighbour correlations, whose pattern is not a cover. The control loses its singlets within $2/J$ at every disorder strength $W$, through the uniform couplings alone; the cover is exact at $W=0$ and decays at $\Gamma\propto W^2$, its curves for all sizes and strengths collapsing onto one function of $t\sqrt\Gamma$ \cite{SM}.

\emph{Cooling into the scar.}---Because $\ket M$ is an eigenstate of every member of the family, it is stationary under any Lindblad dynamics whose Hamiltonian lies in the family and whose jump operators annihilate it. Pumping each dimer into its singlet, $L_{d,m}=\sqrt\gamma\,|\Phi_d\rangle\langle T^m_d|$, an established primitive \cite{Kastoryano2011,Shankar2013,Liu2016,Lin2013,Cole2022,Reiter2016,Mi2024}, is such a channel. One such unit per dimer makes $\ket M$ the unique attractor for every Hamiltonian in the family, static or periodically driven, and since the Hamiltonian never touches the state, its weight can only grow (SM, Propositions~S3 and S4). Unlike in dissipative embeddings of scars \cite{Wang2024,Jiang2025}, the Hamiltonian may be any chaotic member of the family. Figure~\ref{fig:drive}(c) shows a periodic switch between the two Ising Hamiltonians of Fig.~\ref{fig:drive}(a), which heats every other state to the Page entropy within ten periods: pumping at $\gamma=J$ cools the system into the pure area-law scar, each jump resetting one dimer, every trajectory ending exactly in $\ket M$. The approach is a contact process with the scar as its absorbing state \cite{Carollo2019,ODea2024}, the Hamiltonian branching triplets while the pumping removes them. For $\gamma\gtrsim J$ the scar is self-healing: a triplet planted in it rarely avalanches, and in none of our runs at $1.5J$ or above [Fig.~\ref{fig:drive}(d)]. For $\gamma\lesssim J/2$ it usually avalanches into a heated, volume-law state from which the scar nucleates as a rare event, ever rarer with size \cite{SM}.

\emph{Summary and outlook.}---One observation, that a site coupled equally to both ends of a singlet annihilates it, gives the square-lattice Ising model exact eigenstates at every field: products of singlets at the centre of a chaotic spectrum, for every spin and every exchange tensor. Turned into a count of common neighbours, it decides which lattices carry such states; it makes half of all bond disorder harmless, and it makes each one the unique attractor of singlet pumping. These states are exact benchmarks, at any size, for numerical and experimental studies of two-dimensional thermalization, and the other lattices of the classification, bcc and the quasicrystals among them, provide theirs. Multi-spin interactions, far less explored than two-body ones, are the broadest extension: the criterion holds at any order, and the three-spin interaction treated here is its simplest case.

\begin{acknowledgments}
Research supported by a grant from the Fondation Courtois, a Chair of the Institut Courtois, a Discovery Grant from NSERC, and a Canada Research Chair. The authors benefit from their affiliation with the RQMP (\url{https://doi.org/10.69777/309032}). The authors conceived the project and its central ideas. The AI models Claude (Anthropic; Fable 5.1 and 5.2) and Gemini 4 Neon were used, under the authors' direction, to assist with proofs and their cross-checking, with numerical checks and figure preparation. The authors independently checked all results obtained with this assistance and take full responsibility for the content.
\end{acknowledgments}

\clearpage
\onecolumngrid
\begin{center}
{\large\bfseries Supplemental Material}
\end{center}
\setcounter{equation}{0}
\renewcommand{\theequation}{S\arabic{equation}}
\setcounter{figure}{0}
\renewcommand{\thefigure}{S\arabic{figure}}
\setcounter{table}{0}
\renewcommand{\thetable}{S\arabic{table}}
\setcounter{section}{0}
\renewcommand{\thesection}{S\arabic{section}}

\section{The dimer criterion and the wedge form}\label{sec:lemmas}
Lemmas~\ref{thm:dimer} and \ref{prop:wedge} are the dimer criterion and the wedge form used in the main text, both contained in Theorem 2 of Ref.~\cite{Petrovich2024}; the proofs below are self-contained.

\begin{lemma}[dimer criterion]\label{thm:dimer}
$\ket M_s$ is an eigenstate of $H^{zz}=\sum J_{ij}S^z_iS^z_j$ if and only if (a) $W_{dd'}=0$ for all pairs of distinct dimers and (b) for $s\ge1$, no dimer is on a bond. Then $H^{zz}\ket M_s=e\ket M_s$ with $e=-s^2\sum_{d}J_{aa'}$ ($e=0$ for $s\ge1$), and $\ket M_s$ is an exact eigenstate of $H(\Jt,\bh)$ with energy $e\tr\Jt$ for every symmetric $\Jt$ and every $\bh$; if all dimers are off-bond and the bonds are oriented consistently, for every $\Jt$.
\end{lemma}

\begin{proof}
$H^{zz}$ is diagonal and the support of $\ket M_s$ consists of the configurations with $m_{a'}=-m_a$ on every dimer, on which its value is $-\sum_dJ_{aa'}m_d^2+\sum_{d<d'}W_{dd'}m_dm_{d'}$; $\ket M_s$ is an eigenstate iff this is constant there. For $s\ge1$ the monomials $m_d^2$ and $m_dm_{d'}$ are linearly independent functions of $m_d\in\{-s,\dots,s\}$, giving (a) and (b); for $s=\tfrac12$, $m_d^2=\tfrac14$ and only (a) remains. Since $\ket M_s$ is a singlet, $U(g)H^{zz}U(g)^\dagger\ket M_s=e\ket M_s$ for every global rotation $g$, and the rotated Ising couplings $(\bm n\cdot\bS_i)(\bm n\cdot\bS_j)$, whose tensors $\bm n\bm n^{\mathsf T}$ have unit trace, span the symmetric tensors, which gives the energy $e\tr\Jt$ by linearity; the field annihilates a singlet. The antisymmetric part is Lemma~\ref{prop:wedge}.
\end{proof}

\begin{lemma}[wedge form]\label{prop:wedge}
The two-body Hamiltonians $\sum_{i<j}\bS_i^{\mathsf T}\Jt_{ij}\bS_j+\sum_i\bh_i\cdot\bS_i$, with arbitrary bond tensors and site fields, that have $\ket M_s$ as an eigenstate are exactly
\begin{equation}
H=\sum_{d\in M}\big(\mathbf X_d\cdot\bS^+_d+K_d\,\bS_a\cdot\bS_{a'}\big),
\label{eq:wedge}
\end{equation}
where $\mathbf X_d$ is any vector of one-body operators supported outside $d$, $X^\beta_d=h^\beta_d+\sum_{c\notin d}T^{\alpha\beta}_{c,d}S^\alpha_c$, and $K_d\in\mathbb R$ (for $s=\tfrac12$ the intra-dimer term may be any symmetric coupling). The eigenvalue is $-s(s+1)\sum_dK_d$, and $-\tfrac14\sum_d\tr\mathsf K_d$ for symmetric intra-dimer tensors $\mathsf K_d$ at $s=\tfrac12$.
\end{lemma}

\begin{proof}
Every term of \eqref{eq:wedge} annihilates $\ket M_s$ or multiplies it by a constant. Conversely, sort $H\ket M_s$ by the total spin excited on each dimer: spin 1 on one dimer, spin 2 on one dimer (traceless symmetric intra-dimer tensors, absent for $s=\tfrac12$) and spin 1 on two dimers, with coefficients $W^{\alpha\beta}_{dd'}$, the combination $W_{dd'}$ of the tensor components. These components are orthogonal, so each must vanish. The first collects the field difference and the antisymmetric part $\mathbf D_d\cdot(\bS_a\times\bS_{a'})$ of the intra-dimer tensor: on a singlet $(\bS_a\times\bS_{a'})\ket{\Phi_d}=-i\,\bS_a\ket{\Phi_d}$, so it is $(\bh_a-\bh_{a'}-i\mathbf D_d)\cdot\bS_a\ket M_s$, and since $H$ is Hermitian, $\bh$ and $\mathbf D$ are real and both terms vanish. Finally, in the space of couplings between $d$ and $d'$ the four wedges $\bS_b\!\cdot\bS^+_d,\bS_{b'}\!\cdot\bS^+_d,\bS_a\!\cdot\bS^+_{d'},\bS_{a'}\!\cdot\bS^+_{d'}$ span exactly the hyperplane $W_{dd'}=0$.
\end{proof}

\emph{Pair rule.} For uniform couplings $W_{dd'}$ counts bonds with signs, and $W_{dd'}=0$ says that the bonds between two dimers are absent, form a wedge (one site, the apex, adjacent to both ends of the other dimer by its two legs), or are all four present: two wedges, a ring of four bonds with the two dimers as its diagonals, denoted $K_{2,2}$ as in graph theory. A single bond, two disjoint bonds and a path of three bonds are forbidden. Two dimers \emph{touch} when at least one bond joins them. A \emph{cover} below is a pairing of all the sites (a perfect matching), by pairs of any range, obeying this rule; equivalently, the bonds between different dimers can be grouped into wedges. The bonds are oriented \emph{consistently} when the two legs of every wedge point both towards or both away from its apex, as the orientation from one sublattice to the other does on a bipartite lattice. Since $W_{dd'}$ involves only the bonds between $d$ and $d'$, any set of dimers of a cover is a cover of the lattice restricted to its sites: the state remains exact on finite clusters made of whole dimers, with open boundaries. Open rectangles of the square lattice cut dimers of the diagonal VBS, and those searched, from $4\times4$ to $6\times8$, carry no cover.

\section{Which lattices carry covers: proofs of Propositions 2 and 3}\label{sec:class}
\emph{Notation.} Two sites joined by a bond are neighbours, written $a\sim b$; $N(a)$ is the set of neighbours of $a$ and $\deg a=|N(a)|$ its coordination number. A lattice is $k$-regular when every site has $k$ neighbours. A bracket $[\,\cdot\,]$ equals $1$ when the statement inside it holds and $0$ otherwise, so that $[a\sim b]$ counts the bond $ab$. A dimer is on-bond when its two sites are neighbours and off-bond otherwise. For a dimer $d=(a,a')$ let $C(d)=N(a)\cap N(a')$ be its set of middles, and for a site $c$ with partner $c'$ let $W(c)$ be the union of the dimers $d'\ne d(c)$ both of whose ends neighbour $c$ (the dimers $c$ \emph{sees whole}), $m(c)=|W(c)|/2$, and $k(c)$ the number of dimers joined to $d(c)$ by a $K_{2,2}$.

\emph{Proposition S1 (counting identity).} For every site $c$ of a cover,
\begin{equation}
N(c)\setminus\{c'\}=W(c)\cup C(d(c)),\qquad W(c)\cap C(d(c))=\text{the sites of the $K_{2,2}$ partners of }d(c),
\label{eq:decomp}
\end{equation}
hence $\deg c-[c\sim c']=2m(c)+|C(d(c))|-2k(c)$, and summing over the sites,
\begin{equation}
\sum_{d\in M}|C(d)|=\tfrac12\big(\mathcal E-B\big)+2K,
\label{eq:countS}
\end{equation}
with $\mathcal E$ the number of bonds, $B$ the number of on-bond dimers and $K$ the number of $K_{2,2}$ pairs. On a $k$-regular lattice $|C(d)|$ has the parity of $k-[d\text{ on a bond}]$ for every dimer.

\emph{Proof.} Let $x\in N(c)$, $x\neq c'$, with partner $x'$. The bond $cx$ makes the pattern between $d(c)$ and $d(x)$ non-empty, so it is a wedge or a $K_{2,2}$. In a wedge with apex $c$, $c\sim x,x'$ and $x\in W(c)$; in a wedge with apex $x$, $x\sim c,c'$ and $x\in C(d(c))$; in a $K_{2,2}$ both hold. Conversely a site in both sets gives three bonds between the two dimers, which the rule completes to a $K_{2,2}$. This is \eqref{eq:decomp}. Summing its cardinalities: $\sum_c(\deg c-[c\sim c'])=2\mathcal E-2B$; $\sum_cm(c)=\sum_d|C(d)|$, both sides counting the pairs $(c,d)$ in which $c$ sees $d$ whole; $\sum_c|C(d(c))|=2\sum_d|C(d)|$; and $\sum_ck(c)=4K$. Parity follows from the per-site identity. $\square$

Dividing \eqref{eq:countS} by the number $N/2$ of dimers: a dimer has on average at least $(\mathcal E-B)/N$ middles, $k/2-B/N$ on a $k$-regular lattice. This lower bound competes with the geometric upper bound on the number of common neighbours of a pair; the case analysis below compares the two.

\emph{The lattices.} An Archimedean tiling \cite{Grunbaum1987} is a tiling of the plane by regular polygons in which every vertex is surrounded by the same sequence of polygons. Its symbol lists that sequence: $3.4.6.4$ stands for triangle, square, hexagon, square, and $3^6$ for six triangles. The spins sit on the vertices and the bonds are the edges, all of the same length. There are 11 such lattices, listed in Table~\ref{tab:arch} with their names in quantum magnetism \cite{Richter2004,Farnell2018}. A rhombus tiling is a tiling of the plane by rhombi of equal side; the Penrose and Ammann--Beenker quasicrystals and the dice lattice are examples.

\begin{table}[h]
\centering
\begin{tabular}{llcll}
\hline\hline
symbol & name & $k$ & covers & by\\
\hline
$4^4$ & square & 4 & four, off-bond & Prop.~2(i)\\
$3.4.6.4$ & ruby, or bounce (rhombitrihexagonal) & 4 & two, off-bond & S2(b)\\
$3^2.4.3.4$ & Shastry--Sutherland (snub square) & 5 & one, on-bond & S2(c)\\
$3^4.6$ & maple-leaf (snub hexagonal) & 5 & one, on-bond & S2(c)\\
$3^3.4^2$ & trellis & 5 & none & S2(c)\\
$3^6$ & triangular & 6 & none & S2(d)\\
$3.6.3.6$ & kagome & 4 & none & S2(d)\\
$6^3$ & honeycomb & 3 & none & S2(d)\\
$4.8.8$ & square-octagon (CaVO) & 3 & none & S2(d)\\
$3.12.12$ & star & 3 & none & S2(d)\\
$4.6.12$ & square-hexagon-dodecagon (SHD) & 3 & none & S2(d)\\
\hline\hline
\end{tabular}
\caption{The 11 Archimedean lattices: symbol (the polygons around a vertex), name, coordination number $k$, and the covers with the statement that settles each case. The square lattice has two more covers on the $4\times4$ torus.}
\label{tab:arch}
\end{table}

\emph{Proposition S2 (planar tilings).} Let $G$ be an Archimedean lattice or a rhombus tiling, with bonds of unit length. (a) Two sites have at most two common neighbours, and exactly two iff they are opposite corners of a rhombus (or square) face, or the two apices, or the two ends of the shared edge, of two triangles sharing an edge. (b) If $G$ is $4$-regular, every dimer of a cover is an off-bond pair of this kind, $B=K=0$, and every site sees exactly one whole dimer. (c) If $G$ is $5$-regular, every dimer is an edge shared by two triangles, $B=N/2$, $K=0$, and every site sees exactly one whole dimer. (d) If $G$ is $3$-regular and no edge lies in two triangles, or $6$-regular, or $4$-regular without closed loops of four bonds, there is no cover. (e) In a rhombus tiling every dimer of a cover is a diagonal of a tile and every site has four neighbours, outside a vanishing fraction of the sites (with no exception on a finite torus).

\emph{Proof.} (a) The common neighbours of $a,b$ lie on the two unit circles about them, which meet in at most two points; two common neighbours $x,y$ give a closed loop $axby$ of four unit bonds, a rhombus. In an Archimedean tiling the region it encloses, of area at most $1$, is tiled by regular polygons of area at least $\sqrt3/4$: it is a single face or two equilateral triangles, of which $a,b$ are the apices or the ends of the shared edge. In a rhombus tiling the tiles that share parallel edges form chains, the \emph{ribbons}, two of which cross at most once; a ribbon entering the region through one side leaves through the opposite side, so exactly two ribbons cross it and the region is a single tile. (b) Parity (Proposition S1) makes $|C|$ even off-bond and odd on-bond, so $|C|\le2$ off-bond and $|C|\le1$ on-bond, and $\sum_d|C(d)|\le 2(N/2-B)+B=N-B$, while \eqref{eq:countS} gives $N-B/2+2K$; hence $B=K=0$ and $|C(d)|=2$ for all $d$, and the per-site identity reads $4=2m+2$. (c) Now on-bond dimers have $|C|$ even, $\le2$, and off-bond ones odd, $\le1$, so $\sum|C|\le2B+(N/2-B)$ against $\tfrac12(5N/2-B)+2K$: $B=N/2$, $K=0$, $|C(d)|=2$ for all $d$, and $5-1=2m+2$. (d) For $k=3$: on-bond $|C|$ is even, hence $0$ when no edge lies in two triangles, and off-bond $|C|\le1$, so $\sum|C|\le N/2-B$ against $\tfrac34N-\tfrac12B+2K$. For $k=6$: $|C|\le2$ for every dimer by (a), so $\sum|C|\le N$ against $\tfrac12(3N-B)+2K\ge\tfrac54N$, since $B\le N/2$. For $k=4$ without loops of four bonds: $|C|\le1$, so on-bond dimers have $|C|=1$ and off-bond ones $|C|=0$, and $\sum|C|\le B\le N/2$ against $N-\tfrac12B+2K$. (e) A rhombus tiling has $\mathcal E=2N$, and adjacent sites have no common neighbour, since a ribbon entering a triangle could not leave it. Thus $\sum|C|\le2(N/2-B)$ against $N-\tfrac12B+2K$, so $B=K=0$ and $|C(d)|=2$ for every dimer, which is then a diagonal of a tile by (a), and $\deg c=2m(c)+2$. Since $\sum_cm(c)=\sum_d|C(d)|=N$ and $m(c)\ge1$, a site of a rhombus tiling having at least three neighbours, $m(c)=1$ and $\deg c=4$ at every site. On an infinite tiling the sums are densities over growing patches, with negligible boundary terms because a dimer with a middle has range two, and the conclusions hold outside a vanishing fraction of the sites. $\square$

\emph{Proof of Proposition 2.} The identity is Proposition S1. \emph{(i) Square lattice.} By S2(b) every dimer has two middles and every site sees exactly one whole dimer, so the two neighbours of a site that are not middles of its dimer form a dimer. On the torus the pairs with two common neighbours are the plaquette diagonals and, along a side of length four, the wrap-around pairs $(x,y),(x+2,y)$. Let $(0,0)$ pair with $(1,1)$; the middles are $(1,0)$ and $(0,1)$, so $(-1,0)$ and $(0,-1)$ form a dimer, the ``$\backslash$'' diagonal of the plaquette at $(-1,-1)$. The same step at $(-1,0)$, at $(0,-1)$ and at the middle $(1,0)$, whose one whole dimer is already known so that its remaining neighbours are the middles of its own dimer, is forced each time: the partner of one site determines the cover, which is the diagonal VBS. The four choices of the first partner give the four states $\ket{M_{\mu\nu}}$; the pattern has period two, so they close on every even torus and on no torus with an odd side. If instead $(0,0)$ pairs with $(2,0)$ along a side of length four, its middles are $(1,0)$ and $(3,0)$, and $(0,1)$, $(0,-1)$ must form a dimer; they have two common neighbours only if $L_y=4$. Wrap-around pairs therefore occur on the $4\times4$ torus alone, where they give the two extra covers. On a side of length three the rows are triangles and an on-bond dimer has one middle; the count of S2(b) again gives $B=0$, and the same propagation applies.

\emph{(ii) Archimedean lattices.} The triangular lattice $3^6$ is $6$-regular; the honeycomb, square-octagon, star and square-hexagon-dodecagon lattices $6^3$, $4.8.8$, $3.12.12$ and $4.6.12$ are $3$-regular with no edge in two triangles; the kagome lattice $3.6.3.6$ is $4$-regular without loops of four bonds: all are excluded by S2(d). On the trellis lattice $3^3.4^2$ (rows of squares alternating with rows of triangles) S2(c) puts every dimer on a zigzag edge $ab$ shared by two triangles; its middles are the other zigzag neighbour and one horizontal neighbour of $a$, so the remaining neighbours of $a$, one horizontal and one across a square, would have to form a dimer, which is off-bond, contradicting $B=N/2$. The ruby lattice $3.4.6.4$ is $4$-regular and its pairs with two common neighbours are the diagonals of its squares, so by S2(b) a cover is a choice of one diagonal per square covering every site, which lies in two squares, exactly once. Around a hexagon $h_1\cdots h_6$ the six squares on its edges $h_ih_{i+1}$ must cover each of $h_1,\dots,h_6$ exactly once, so every square covers $h_i$ or every square covers $h_{i+1}$: a sense of rotation per hexagon, which two hexagons sharing a square must share. Hence exactly two covers, mirror images of each other; every site sees whole the diagonal of its other square, so every touching pair is a wedge. Each wedge lies in one square, so orienting the four bonds of every square from one diagonal to the other orients the bonds consistently, for both covers, although the lattice is not bipartite. $3^2.4.3.4$ is the Shastry--Sutherland lattice with equal couplings: it is $5$-regular, each site has exactly one edge shared by two triangles, and these edges pair all the sites, the only candidate by S2(c), which is the Shastry--Sutherland dimer state \cite{Shastry1981,McClarty2020}; the two remaining neighbours of a site lie on such an edge, so the rule holds. $3^4.6$ is the maple-leaf lattice of Ref.~\cite{McClarty2020}, the triangular lattice with a $\sqrt7\times\sqrt7$ superlattice of sites removed; every site $a$ has one removed neighbour, in direction $\bm\delta_a$, and three edges shared by two triangles, to its neighbours at $\pm120^\circ$ and $180^\circ$ from $\bm\delta_a$, the candidates of S2(c). Pairing $a$ along the third leaves its neighbours at $\pm60^\circ$, which are not adjacent, to form the dimer that $a$ sees whole, contradicting $B=N/2$. The rule $a\mapsto a+R_{-120^\circ}\bm\delta_a$, with $R_\theta$ the rotation by $\theta$, is its own inverse and gives the known dimer state; the mirror rule is not, and the choice at one site propagates, so this cover is the only one.

\emph{(iii) Three dimensions.} bcc is $8$-regular; its pairs have $4$ common neighbours at the second-neighbour distance $(1,0,0)$, $2$ at $(1,1,0)$, $1$ at $(1,1,1)$, and adjacent sites have none, so parity forbids on-bond dimers and \eqref{eq:countS} forces every dimer onto a $(1,0,0)$ pair with $K=0$. Such covers exist on $L^3$ cubic cells with even $L\ge4$: with the sites at the points $(x,y,z)$ whose coordinates are all even or all odd, pair $(x,y,z)$ with $(x,y,z+2)$ on the even sublattice and with $(x,y+2,z)$ on the odd one whenever $(y-z)/2$ is even; every touching pair is a wedge (checked for $L=4,6,8$, $N\le1024$). Simple cubic ($k=6$): off-bond pairs have $|C|\le2$ and on-bond ones $|C|=0$ against the odd parity required, so $\sum|C|\le N$ against $\tfrac32N$. fcc and hcp ($k=12$): $|C|\le4$ against an average of at least $11/2$. Diamond ($k=4$): adjacent sites have no common neighbour against the odd parity required, and off-bond pairs have $|C|\le1$ against even parity, so $\sum|C|=0$ against $N$. Pyrochlore ($k=6$): on-bond dimers have $|C|=2$ against odd parity, off-bond pairs $|C|\le1$ against even parity.

\emph{(iv) Rhombus tilings.} This is S2(e): in the Penrose and Ammann--Beenker tilings and in the dice lattice a nonzero fraction of the sites have three neighbours. $\square$

\emph{Proof of Proposition 3.} De Bruijn's construction \cite{deBruijn1981} starts from a grid of $n\ge3$ families of parallel lines, two lines of different families meeting once and no three lines through one point (a non-singular multigrid), and takes its dual: one vertex of the tiling for every cell of the grid, one edge of unit length for every segment between two crossings, perpendicular to it, and one rhombus for every crossing. Five families of equidistant lines at angles of $72^\circ$ give the Penrose tilings and their generalizations, four at $45^\circ$ the Ammann--Beenker tiling. Each rhombus is thus the crossing of two lines $\ell,\ell'$ and has two edges for each of them; the rhombi along a line $\ell$ form its \emph{ribbon}, consecutive ones sharing an edge, and every edge belongs to exactly one ribbon. The lattice of the spins, the \emph{medial} lattice of the tiling, has the edges of the tiling as sites, two of them neighbours iff they are adjacent sides of a tile; it is $4$-regular and planar. Let each rhombus $F=\ell\cap\ell'$ \emph{claim} one of its two lines and pair the midpoints of its two edges of the claimed direction. This pairs all the sites iff every edge is claimed by exactly one of its two rhombi, i.e.\ iff along every line the claims alternate. With $n_\ell(F)$ the position of $F$ along $\ell$ and one unknown phase $\varphi_\ell$ per line, alternation is the system $\varphi_\ell+\varphi_{\ell'}\equiv1+n_\ell(F)+n_{\ell'}(F)\pmod2$, one equation per intersection. Around a closed cycle of lines the phases cancel and the condition for solvability is that the number of lines crossing the sides of the polygon of intersections, strictly between its corners, be even. For a triangle of three lines this holds, since every other line crosses a closed curve an even number of times and never at a corner; and every cycle is a sum of triangles, because two lines of different families always meet (a cycle through two parallel lines is split by a line of a third family). The system is therefore solvable, with exactly two solutions, $\varphi\to\varphi+1$. For the pair rule, let $e$ be an edge with tiles $F$ (claiming it) and $F'$, and $\hat e$ its partner, the opposite edge of $F$: the common neighbours of $e$ and $\hat e$ are the other two edges of $F$, and the other two neighbours of $e$ are the two edges of $F'$ adjacent to it, which are exactly the pair claimed by $F'$. So $|C|=2$, $m=1$, $K=0$, and every touching pair is a wedge, which lies in a single tile; orienting the four bonds of every tile from one pair of opposite edges to the other orients them consistently. For the square lattice, which is its own medial lattice, the ribbons are the rows and columns of plaquettes and the construction is the diagonal VBS rotated by $45^\circ$. $\square$

\emph{Verification.} The pair rule was searched exhaustively, over pairings of any range, by backtracking (at each step the uncovered site with the fewest compatible partners; the $8\times8$ torus, with $63!!\approx10^{44}$ pairings, takes one second) on tori of the 11 Archimedean lattices: six covers on the $4\times4$ square torus, four on the even tori from $4\times6$ to $8\times8$ and none on nine tori with an odd side, from $3\times4$ to $6\times7$; two on the ruby lattice (24 and 54 sites), one each on the Shastry--Sutherland (16 and 36 sites) and maple-leaf lattices (42 and 168 sites), none on the other seven (12 to 54 sites). On the icosahedron it finds five covers, all on edges, which together use each of the 30 edges once; for $s=\tfrac12$ they span the level $E=-6$ of $\sum\sigma^z_i\sigma^z_j+h\sum\sigma^x_i+g\sum\sigma^z_i$, five-fold at each of four fields examined, $g=0$ included. The construction on the Penrose tiling with spins on its edges was verified on two patches of $1190$ and $1181$ rhombi: the alternation propagated without conflict and all $1839$ and $1822$ pairs of touching dimers whose neighbourhoods lie inside the patch were wedges.

\section{Multi-spin interactions: the criterion and the proof of Proposition 4}\label{sec:multi}
Give the two sites of every dimer the signs $\tau_a=+1$, $\tau_{a'}=-1$. For a set $\he$ of sites let $T(\he)$ be the set of dimers that $\he$ meets exactly once and $\tau(\he)=\prod_{i\in\he}\tau_i$. For interactions $\sum_{\he}J_\he\prod_{i\in\he}\sigma^z_i$ of order $k$ define $c^{(k)}_T=\sum_{\he:\,T(\he)=T}J_\he\tau(\he)$.

\begin{lemma}[multi-spin criterion]\label{thm:kbody}
(i) $\ket M_{1/2}$ is an eigenstate of $\sum_{\he}J_\he\prod_{i\in\he}\sigma^z_i$, the sum over sets of $k$ sites, iff $c^{(k)}_T=0$ for every non-empty $T$; the eigenvalue is $c^{(k)}_\varnothing$, which vanishes when $k$ is odd. (ii) It is then an exact eigenstate of $\sum_{\he}J_\he\Tt_k^{\alpha_1\cdots\alpha_k}S^{\alpha_1}_{\he_1}\cdots S^{\alpha_k}_{\he_k}+\bh\cdot\bS_{\rm tot}$ for every tensor $\Tt_k$ symmetric under permutation of its indices and every field.
\end{lemma}

\begin{proof}
(i) On the antiparallel configurations $\prod_{i\in\he}\sigma^z_i=\tau(\he)\prod_{d\in T(\he)}\epsilon_d$ with $\epsilon_d=\pm1$ the orientation of dimer $d$, since a dimer met twice contributes $-1$; distinct monomials in the $\epsilon_d$ are linearly independent, and an odd set cannot meet every dimer twice. (ii) The products $\prod_{i\in\he}(\bm n\cdot\bS_i)$ span, over unit vectors $\bm n$, the symmetric tensors, so the family is the span of the global rotations of the Ising interaction, of which the singlet $\ket M$ is an eigenstate with the same eigenvalue.
\end{proof}

\emph{Proof of Proposition 4.} For a site $\bm r$ of the diagonal VBS let $\bm r'$ be its partner, $c_1,c_2$ the middles of their dimer and $(u,v)$ the dimer that $\bm r$ sees whole, so that $N(\bm r)=\{c_1,c_2,u,v\}$ and $N(\bm r')=\{c_1,c_2,u',v'\}$. Write $t(\bm r;x,y)=\Tt_3^{\alpha\beta\gamma}S^\alpha_{\bm r}S^\beta_xS^\gamma_y$, which is symmetric in $x,y$ because $\Tt_3$ is symmetric in its last two indices. The six triples at $\bm r$ and the six at $\bm r'$ combine into terms that annihilate $\ket M$: (i) $t(\bm r;c_i,u)+t(\bm r;c_i,v)$, $i=1,2$, contains $\bS_u+\bS_v$; (ii) $t(\bm r;c_1,c_2)+t(\bm r';c_1,c_2)$ contains $\bS_{\bm r}+\bS_{\bm r'}$; (iii) on the singlet $(u,v)$, $S^\beta_uS^\gamma_v=-S^\beta_uS^\gamma_u$, whose symmetric part is $-\tfrac14\delta^{\beta\gamma}$ for $s=\tfrac12$, so $t(\bm r;u,v)\ket M=-\tfrac14\Tt_3^{\alpha\beta\beta}S^\alpha_{\bm r}\ket M$, and with the same term at $\bm r'$ this is proportional to $\Tt_3^{\alpha\beta\beta}(S^\alpha_{\bm r}+S^\alpha_{\bm r'})\ket M=0$. Hence $H_3\ket M=0$. For $s\ge1$ the symmetric part of $S^\beta_uS^\gamma_u$ has a traceless piece, a spin-2 excitation of the dimer, and (iii) holds only for $\Tt_3^{\alpha\beta\gamma}=v^\alpha\delta^{\beta\gamma}$: the interaction $(\bm v\cdot\bS_{\bm r})(\bS_n\cdot\bS_{n'})$ keeps the state exact at every spin.

The argument uses only that every neighbour of a site is a middle of its dimer or belongs to a dimer that the site sees whole, as many of these at both ends of a dimer. By Eq.~(\ref{eq:decomp}) this holds for every cover without on-bond dimers and $K_{2,2}$ pairs on a $k$-regular lattice, so the three-spin interaction of every site with every pair of its neighbours also annihilates the covers of the ruby and bcc lattices and of the rhombus tilings with spins on their edges (for bcc, where a site has four middles and sees two dimers whole, checked through the coefficients of Lemma~\ref{thm:kbody} for $L=4,6$). On the square lattice each pair (i) contains one straight and one corner triple, which is why neither kind alone is admissible. In the Ising form the coefficients $c^{(3)}_T$ of Lemma~\ref{thm:kbody} vanish accordingly, as do those of the five-spin stars $\prod_{i\in\{\bm r\}\cup N(\bm r)}\sigma^z_i$; evaluated on $L\times L$ tori, $L=6,8,10$, the coefficients show that the VBS is not an eigenstate of the straight triples alone, the corner triples alone, the plaquette four-spin term, or second- and third-neighbour exchange. On $4\times4$ the residual of $H_3\ket{M_{00}}$ in the full Hilbert space is $10^{-16}$ for a random $\Tt_3$ symmetric in its last two indices and of order one for a generic one. Theorem 1 of Ref.~\cite{Petrovich2024} tests a product state against Hamiltonians bilinear in the operators of its factors; a triple meeting three dimers is trilinear, and becomes bilinear only for factors as large as strips of the lattice, for which the test restates the question.

\section{Scar status: level statistics and the zero-energy space}\label{sec:scar}
Model: $H=\sum_{\langle ij\rangle}\sigma^z_i\sigma^z_j+h\sum_i\sigma^x_i+g\sum_i\sigma^z_i$ on the $4\times4$ torus at $(h,g)=(3,\pm1)$ and $(1,0.5)$, and on the 20-site torus with vectors $(4,2),(-2,4)$ at $(3,1)$; energies in units of the coupling, entropies in nats for half of the torus.

\emph{Symmetry resolution and level statistics.} The bonds of the $4\times4$ torus are those of the four-dimensional hypercube, whose symmetry group, the permutations of the sites that preserve the bonds, has order $384$, three times the order of the space group; the Ising family inherits all of it, so a momentum sector resolved by its little group is still reducible. The irreducible blocks with one-dimensional characters are those of the group generated by the translations and the two reflections $x\to-x$, $y\to-y$, at momenta $k\in\{0,\pi\}^2$ with reflection parities obeying $p_xp_y\cos(k_x+k_y)=-1$: four distinct spectra of $813$ to $1353$ levels. Over the middle half of each, $\langle r\rangle=0.508$--$0.539$ at $(3,\pm1)$ and $0.514$--$0.542$ at $(1,0.5)$, pooled $0.524\pm0.006$ and $0.531\pm0.006$ over $2032$ ratios (GOE $0.5307$, Poisson $0.3863$). Momentum sectors resolved by one mirror give $0.41$--$0.47$; each is, to $10^{-13}$, the union of two irreducible spectra, and two superposed GOE sequences give $0.42$ for equal sizes \cite{Giraud2022} and $0.43$ for the size ratio $1353:813$.

\emph{Eigenstates.} All $65\,536$ eigenstates were computed block by block. At $(3,\pm1)$ the $7481$ eigenstates with $0<|E|<2$ have $S=4.82$ on average (minimum $3.88$; Page value $5.045$) and a correlator $\langle\bS_{(0,0)}\!\cdot\!\bS_{(1,1)}\rangle=-0.017\pm0.017$ on a dimer of $\ket{M_{00}}$, where the cover has $S=4\ln2=2.773$ and $-3/4$. Exactly six eigenvalues vanish ($|E|<3\times10^{-15}$; the next level is at $7.6\times10^{-4}$, of the order of the mean level spacing), and their eigenvectors span the six covers (residuals below $10^{-12}$), whose pairwise overlaps are $1/16$; the covers are the least entangled states of that space. The same holds at $(1,0.5)$.

\emph{Hidden, then isolated.} At $g=0$, $h=3$, the zero-energy space is $770$-fold degenerate and contains the six covers, and the eigenvectors returned by the diagonalization have $S$ between $4.02$ and $4.96$. The longitudinal field has no matrix elements within that space (Frobenius norm $4\times10^{-12}$ of the projected operator), so the splitting is of second order: at $g=10^{-4}$, $10^{-3}$, $10^{-2}$ the $764$ other zero modes lie at $|E|=c\,g^2$ with $c$ between $10^{-3}$ and $10^2$, while the covers, which the field annihilates, stay at $|E|<3\times10^{-15}$ for every $g$ up to $2$.

\emph{Tilted 20-site torus.} The $5982$ eigenstates nearest $E=0$ in the twenty momentum sectors ($|E|<0.11$, shift-invert Lanczos, residuals below $2\times10^{-10}$) have $S$ between $6.31$ and $6.36$ (Page value $6.431$); the only eigenvalues below $10^{-8}$ are the four cover states, with $S=4\ln2$ for each cover.

\section{Disorder: the short-time law and the quench of Fig.~2(b)}\label{sec:driven}
\emph{Short-time law.} For two-body deviations $V=\sum_{i<j}\bS_i^{\mathsf T}\delta\Jt_{ij}\bS_j+\sum_i\delta\bh_i\cdot\bS_i$ with no coupling inside a dimer, as for any bond disorder on an off-bond cover, and $s=\tfrac12$, the sorting of Lemma~\ref{prop:wedge} gives $(V-\langle V\rangle)\ket M=\sum_d\tfrac12(\delta\bh_a-\delta\bh_{a'})\cdot(\bS_a-\bS_{a'})\ket M+\sum_{d<d'}\sum_{\alpha\beta}\delta W^{\alpha\beta}_{dd'}S^\alpha_aS^\beta_b\ket M$, with $\delta\Wt_{dd'}=\delta\Jt_{ab}-\delta\Jt_{ab'}-\delta\Jt_{a'b}+\delta\Jt_{a'b'}$, every tensor written from $d$ to $d'$ ($\delta\Jt_{ji}=\delta\Jt_{ij}^{\mathsf T}$). The states $(S^\alpha_a-S^\alpha_{a'})\ket M$, of norm $1$, and $S^\alpha_aS^\beta_b\ket M$, of norm $\tfrac14$, are mutually orthogonal, so the triplet weight $p_d=\langle P^T_d\rangle$ of a dimer obeys
\begin{equation}
p_d(t)=\Gamma_dt^2+O(t^4),\qquad\Gamma_d=\|P^T_dV\ket M\|^2=\tfrac14|\delta\bh_a-\delta\bh_{a'}|^2+\tfrac1{16}\sum_{d'\ne d}\|\delta\Wt_{dd'}\|_F^2,
\label{eq:gamma_general}
\end{equation}
the $t^3$ term vanishing for real Hamiltonians. A coupling inside a dimer changes $\Gamma_d$ only through its antisymmetric part $\delta\mathbf D_d\cdot(\bS_a\times\bS_{a'})$, which adds $\tfrac14|\delta\mathbf D_d|^2$. For a cover the uniform part of $H$ contributes nothing; for a singlet product that is not a cover the formula holds with the full couplings in place of the deviations.

\begin{figure}[h]
\centering
\includegraphics[width=0.5\textwidth]{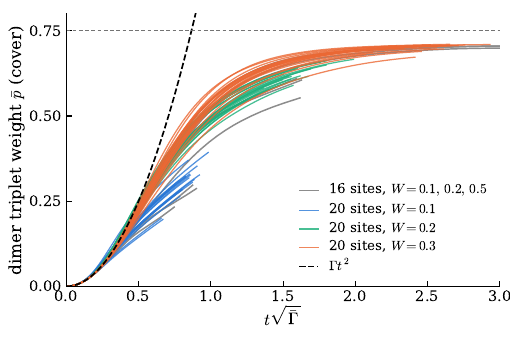}
\caption{Mean dimer triplet weight of the cover against $t\sqrt{\bar\Gamma}$, with $\bar\Gamma$ the dimer average of Eq.~(\ref{eq:gamma_general}) for each realization: 16 sites (grey, three realizations per strength) and 20 sites (20 of the 100 realizations per strength); dashed: $\Gamma t^2$.}
\label{fig:disorder}
\end{figure}

\emph{Quench.} Model: $H=\sum_{\langle ij\rangle}J_{ij}\bS_i\cdot\bS_j+\sum_ih_iS^z_i$ with $J_{ij}=J(1+W\epsilon_{ij})$ on every bond and $h_i=WJ\eta_i$ on every site, $\epsilon_{ij},\eta_i$ independent and uniform in $[-1,1]$; nothing is balanced. Initial states: the cover $\ket{M_{00}}$ and the non-cover $\ket C$ with singlets on $(x,y)$--$(x+1,y+1)$ for every site with $x$ even. Both are products of next-nearest-neighbour singlets, with zero energy and vanishing nearest-neighbour correlators; in $\ket C$ the uniform couplings already give each dimer $W_{dd'}=\pm1$ with four other dimers, so that $\Gamma=\tfrac3{16}\sum_{d'}W_{dd'}^2=\tfrac34$ at $W=0$ by Eq.~(\ref{eq:gamma_general}); on the 20-site torus two of the four coincide, $W_{dd'}=(-2,1,1)$ and $\Gamma=\tfrac98$. The cover has $\Gamma=0$. On 20 sites, $100$ realizations at each of $W=0.1$, $0.2$, $0.3$: the dimer-averaged $\bar\Gamma$ of the cover is $0.0068$, $0.026$, $0.061$, against the disorder average $\tfrac23W^2J^2$ of Eq.~(\ref{eq:gamma_general}); $p_d(t)/\Gamma_dt^2$ lies between $0.9995$ and $1.0008$ at $t=0.1/J$ in all $300$ quenches; the cover keeps $95\%$ of its singlet weight until $t=2.7/J$, $1.4/J$ and $1.0/J$, while the non-cover is half-way to infinite temperature at $0.7/J$ for every $W$. Against $t\sqrt{\bar\Gamma}$ the cover curves of both sizes and all strengths fall on one curve (Fig.~\ref{fig:disorder}). Method: Chebyshev propagation in the sector $S^z_{\rm tot}=0$, checked against direct integration in the full space ($10^{-14}$).

\section{Cooling: harmless channels, the unique attractor and the rates}\label{sec:pump}
\emph{Channels.} For $\mathcal L\rho=-i[H,\rho]+\sum_k(L_k\rho L_k^\dagger-\tfrac12\{L_k^\dagger L_k,\rho\})$ the state $\ket M\bra M$ is stationary if $H\ket M\propto\ket M$ and each $L_k$ annihilates $\ket M$ or is Hermitian with $\ket M$ as an eigenvector. By the sorting argument of Lemma~\ref{prop:wedge} without its last step, which uses Hermiticity, the one- and two-body operators annihilating $\ket M$ are sums of wedges $\mathbf X_d\cdot\bS^+_d$ and of operators on a single dimer that annihilate its singlet, the resets below among them. The harmless channels are thus the pumping itself, the collective ones, $L=S^\alpha_{\rm tot}$, the dimer-collective ones, $L=S^{+\alpha}_d$ \cite{Zanardi1997,Lidar1998}, and wedge-correlated ones; every single-site channel is harmful. The resets $L_{d,m}=\sqrt\gamma\,|\Phi_d\rangle\langle T^m_d|$, $m=0,\pm1$, annihilate $\ket M$ and satisfy $\sum_mL^\dagger_{d,m}L_{d,m}=\gamma P^T_d$.

\emph{Proposition S3 (unique attractor).} With the resets on every dimer of a singlet cover and any Hamiltonian $H(t)$, static or periodic, the dynamics has a unique stationary state, to which every initial state converges; it is $\ket M\bra M$ if $H(t)\ket M\propto\ket M$ for all $t$.

\emph{Proof.} The support $S$ of a stationary state $\rho$, the span of its eigenvectors with non-zero weight, is invariant under the Kraus operators $K$ of the map $\Phi$ over one period $\tau$: $\Phi(\rho)=\rho$ is a sum, and an integral over the jump times with a continuous integrand, of positive terms $K\rho K^\dagger$, none of which can have weight outside $S$. The Dyson series gives a Kraus family: the no-jump operator $K_0=\mathcal T\exp(-i\int_0^\tau H_{\rm eff}\,dt)$, $H_{\rm eff}=H-\tfrac i2\sum_kL_k^\dagger L_k$, which is invertible so that $K_0S=S$, and for a jump at time $t$ the operator $U_{\rm eff}(\tau,t)L_{d,m}U_{\rm eff}(t,0)$, which tends to $L_{d,m}K_0$ as $t\to\tau$; hence $L_{d,m}S\subseteq S$. The resets of different dimers commute, and for $\psi\in S$ and a set $A$ of dimers with triplet labels $m_d$, $\psi_A=\prod_{d\in A}L_{d,m_d}\psi\propto|\Phi_A\rangle\otimes\chi_A\in S$, with $\chi_A=\langle T^{m}_A|\psi\rangle$ a vector on the remaining dimers. Suppose $\ket M\notin S$ and $\psi\neq0$. For $A$ the set of all dimers $\psi_A\propto\ket M$, so $\chi_A=0$. If $\chi_{A\cup d}=0$ for all $d\notin A$ and all labels, then $\chi_A$ has no triplet on any of its dimers, so $\psi_A\propto\ket M$ and hence $\chi_A=0$. Descending to $|A|=1$ gives $\langle T^m_d|\psi\rangle=0$ for all $d,m$, i.e.\ $\psi\propto\ket M$, a contradiction. Hence every nonzero subspace invariant under $K_0$ and the resets contains $\ket M$. Two stationary states would have orthogonal supports, so the stationary state is unique. The Kraus family of $\Phi^p$ contains $K_0^p$ and $L_{d,m}K_0^p$, so the same holds for every power of $\Phi$: the support does not split into subspaces permuted cyclically, the peripheral spectrum of $\Phi$, its eigenvalues of modulus one, is $\{1\}$ \cite{Baumgartner2008}, and $\Phi^n(\rho)$ converges for every $\rho$. For a static $H$ the argument applies to $e^{\tau\mathcal L}$ for any $\tau$. $\square$

\emph{Proposition S4 (monotone weight and the bound).} For $H(t)$ in the family and resets only,
\begin{equation}
\partial_t\langle M|\rho|M\rangle=\gamma\sum_{d,m}\langle T^m_d\otimes M_{\bar d}|\rho|T^m_d\otimes M_{\bar d}\rangle\ge0 ,
\end{equation}
with $\ket{M_{\bar d}}$ the cover without the dimer $d$: $H\ket M\propto\ket M$ removes the commutator, $P^T_d\ket M=0$ the anticommutator, and $L^\dagger_{d,m}\ket M=\sqrt\gamma\,|T^m_d\otimes M_{\bar d}\rangle$. The weight $F$ of the scar grows only through the one-triplet manifold, of weight $w_1\le1-F$, so $\partial_t(1-F)\ge-\gamma(1-F)$: the cooling rate $r=\gamma\,\overline{w_1}/(1-F)$ never exceeds the pumping rate.

\emph{Protocol and rates.} The Hamiltonian alternates between the mixed-field Ising models $(h,g)=(1,0.5)$ and $(3,-1)$ for a time $1/J$ each, which heats every state other than the covers to infinite temperature within ten periods; the resets act on the dimers of $\ket{M_{00}}$. Quantum trajectories (no-jump evolution by a Faber expansion, checked against exact Lindblad evolution on the $4\times2$ torus with its rungs counted twice), $100$ to $400$ per point, on the $4\times4$, 20-site and $4\times6$ tori. Table~\ref{tab:pump} gives the late-time rate of $1-\overline F$ and the probability that one triplet planted in the scar grows to at least half of the dimers. For $\gamma\ge J$ the rates of 16 and 20 sites agree, and the $4\times6$ torus at $\gamma=3J$ reaches the exact scar within two periods; at $\gamma=0.3J$ the system first heats to a quasi-stationary state with $0.57$ triplets per dimer and escapes at the rate $\gamma\overline{w_1}$ of Proposition S4, which falls with size: the hot state lives about $90$ periods on 16 sites and $260$ on 20. Single-site decay at rate $\kappa$ per site alone gives $F\ge e^{-\kappa Nt/2}$, an equality up to the return of trajectories that have jumped, because $\sum_j\sigma^+_j\sigma^-_j\ket M=\tfrac N2\ket M$; for $\kappa=0.02J$ on 16 sites this is $0.008$ after fifteen periods, whereas pumping at $\gamma=3J$ holds a steady fidelity $0.93$.

\begin{table}[h]
\centering
\begin{tabular}{lcccccc}
\hline\hline
$\gamma/J$ & 0.3 & 0.5 & 1 & 2 & 3 & 10\\
\hline
rate $r/J$, $N=16$ & 0.0055 & 0.028 & 0.19 & 0.77 & $\ge1.35$ & $\ge8.8$\\
rate $r/J$, $N=20$ & 0.0019 & -- & 0.19 & $\ge0.75$ & $\ge1.35$ & $\ge9.4$\\
avalanche probability, $N=16$ & -- & 0.67 & 0.02 & 0 & 0 & 0\\
avalanche probability, $N=20$ & -- & 0.67 & 0 & 0 & -- & --\\
\hline\hline
\end{tabular}
\caption{Cooling rate of $1-\overline F$ per unit time (lower bounds where the scar is reached within the time resolution; the exact bound $r\le\gamma$ is approached only for $\gamma\gg J$) and probability that one planted triplet avalanches, for the protocol of Fig.~2(c,d).}
\label{tab:pump}
\end{table}

\end{document}